\documentclass{aa}  

\usepackage{graphicx}
\usepackage{color}
\usepackage{appendix}
\usepackage{txfonts}
\providecommand{\teff}{\ensuremath{T_{\rm eff}}}
\providecommand{\numax}{\ensuremath{\nu_{\rm max}}}
\providecommand{\deltanu}{\ensuremath{\langle \Delta\nu \rangle}}
\providecommand{\fbol}{\ensuremath{F_{\rm bol}}}
\providecommand{\lum}{\ensuremath{L_{\star}}}
\providecommand{\rad}{\ensuremath{R_{\star}}}

\providecommand{\logg}{\ensuremath{\log g}}
\providecommand{\msol}{${\cal M}_{\odot}$}
\providecommand{\rsol}{${\cal R}_{\odot}$}
\providecommand{\lsol}{${\cal L}_{\odot}$}
\newcommand{\refiva}{K18}
\newcommand{\refcre}{C12}

\begin{document}

  \title{First detection of oscillations in the Halo giant \object{HD 122563}: \\
validation of seismic scaling relations and new parameters}
\titlerunning{Asteroseismic Inference on HD\,122563}

   \author{O. Creevey
          \inst{1}\fnmsep\thanks{ocreevey@oca.eu}
          \and
          F. Grundahl\inst{2}
          \and 
          F. Th\'evenin\inst{1}
          \and
          E. Corsaro\inst{3}
          \and 
          P. L. Pall\'e\inst{4}
          \and 
          D. Salabert\inst{5,6}\fnmsep\thanks{Previous affiliation}
		  \and \\
          B. Pichon\inst{1}
          \and
          R. Collet\inst{2}
          \and
         L. Bigot\inst{1}
           \and 
          V. Antoci\inst{2}
          \and 
          M. F. Andersen\inst{2}
          }

   \institute{Université Côte d'Azur, Observatoire de la Côte d'Azur, 
   				CNRS, Laboratoire Lagrange, France\\
              \email{ocreevey@oca.eu}
         \and Stellar Astrophysics Centre, Department of Physics and Astronomy, Aarhus University, Ny Munkegade 120, 8000 Aarhus C, Denmark
         \and INAF - Osservatorio Astrofisico di Catania, via S. Sofia 78, 95123-I, Catania, Italy
         \and Instituto de Astrofisica de Canarias, 38205 La Laguna, Tenerife, Spain
         \and IRFU, CEA, Université Paris-Saclay, F-91191, Gif-sur-Yvette, France
 \and  Universit\'e Paris Diderot, AIM, Sorbonne Paris Cit\'e, CEA, CNRS, F-91191 Gif-sur-Yvette, France
             }

   \date{Received November 26, 2018; accepted November 26, 2018}

 
  \abstract
  {}
   {The nearby metal-poor giant \object{HD\,122563} is an important astrophysical laboratory for 
   which to test stellar atmospheric and interior physics.  It is also a benchmark star for
   which to calibrate methods to apply to large scale surveys.  Recently it has been 
   remeasured using various methodologies given the new high precision instruments at our
   disposal.  However, inconsistencies in the observations and models have been found.
   In order to better characterise this star using complementary techniques we have been 
   measuring its radial velocities since 2016 using the Hertzsprung telescope (SONG network node) in order to detect oscillations.} 
   {}
   {In this work we report the first detections
   of sun-like oscillations in this star, and to our knowledge, a detection in the 
   most metal-poor giant to date.  We apply the classical seismic scaling relation to derive a new surface gravity
   for HD\,122563 of {\logg$_{\nu} = 1.39 \pm 0.01$ dex. } 
   Reasonable constraints on the mass imposed by its PopII giant classification then yields a radius of $30.8 \pm 1.0$ \rsol.
   By coupling this new radius with recent interferometric measurements we infer a distance to the star of {306 $\pm$ 9} pc, which places it further away
   than previously thought and inconsistent with the Hipparcos parallax.
   Independent data from the Gaia mission corroborates the distance hypothesis ($d_{\rm GDR2}$ = 290 $\pm$ 5 pc), and thus the updated fundamental parameters. }
   {We confirm the validity of the seismic scaling relation {without corrections} for surface gravity in 
   metal-poor and evolved star regimes.  The remaining discrepancy of 0.04 dex between \logg$_{\rm GDR2}$ ($= 1.43 \pm 0.03$) reduces to 0.02 dex by applying corrections to the scaling relations based on the mean molecular weight and adiabatic exponent. 
    The new constraints on the HR diagram (\lum$_{\nu} = 381 \pm 26$) significantly reduce the disagreement between the stellar parameters and evolution models, however, a discrepancy on the order of 150 K still exists. Fine-tuned stellar evolution calculations show that this can be reconciled by changing the mixing-length parameter by an amount (--0.35) that is in agreement with predictions from recent 3D simulations and empirical results.
   Asteroseismic measurements are continuing, and analysis of the full frequency data complemented by a distance estimate promises to bring important constraints on our understanding of this star and of the accurate calibration of the seismic scaling relations in this regime.}
   
   {}


   \maketitle
%


\section{Introduction}
\label{introduction}
\object{HD\,122563} ($V$=6.2 mag, 14$^{\rm h}$02$^{\rm m}$~31.8$^{\rm s}$, +09$^{\circ}$41${'}$09.95${"}$) is {one of the brightest and closest metal-poor [M/H]~=~--2.4 \citep{collet18,pra17} giant stars that offers the possibility to be observed and analysed using many different methodologies.   There are few stars for which such a complete set of observations can be obtained.  The advantage of this is that we can obtain robust (almost) model-independent determinations of many of its fundamental parameters, but many of these can also be compared by using different methodologies, and so it allows us to investigate sources of systematic errors. }
For example, 
	it serves as a benchmark star for testing 
    stellar astrophysics, such as, non-LTE effects 
	in stellar atmospheres \citep{heiter15,the99,collet05}, 
    or 3D stellar atmosphere structure 
    \citep{pra17, collet18}.
    In stellar evolution models strong assumptions on its age, mass and initial helium abundance
    can be made which allows one to investigate tunable parameters or details of the interior physics.
    HD\,122563 also contributes to calibrating large Galactic 
    surveys, e.g. \cite{gilmore2012}, which
	aim to understand the evolution of the Milky Way \citep{jofre17}.
	
	In \citet{cre12b}, from hereon \refcre, strong constraints were placed on the position of this star 
	in the Hertzsprung-Russell diagram, in particular with the determination of its interferometric diameter. 
    It was found that classical stellar evolution models
	were unable to reproduce its position within the error box.   One of the 
	proposed solutions by these authors was to change the mixing-length parameter $\alpha$ 
	in the stellar evolution codes, and they managed to produce a model which
	correctly fit the observational data. Such changes are not unrealistic e.g. \citep{bonaca2012,creevey2017,tayar2017, joyce18, viani18},
	however, the size of the modification suggested that something should be addressed
	either in the models for very metal-poor stars or in the observations.
    Even with such constraints on the
	model parameters, its derived age was not well constrained, 
	due to degeneracies between the unobservable mass and initial 
	helium mass fraction, see \cite{lebreton1999} for a discussion.  
    \refcre\ proposed that asteroseismic 
	observations could help to constrain these other parameters, which 
	would improve the age determination and perhaps shed some light on the 
	difficulty of matching the HR diagram constraints with classical models.
    
	HD\,122563 has also been the subject of several recent atmospheric studies \citep{amarsi18}, 
	and some discrepancies in analyses were noted, for example, the 3D non-LTE analysis of hydrogen lines
    shows a spread of \teff\ values depending on the line used.   
	\citet{collet18} suggested that a lowering of the 
    surface gravity from their adopted $\log g = 1.61 \pm 0.07$ 
    would reduce the oxygen abundance discrepancy between 
    molecular and atomic  species.  C12 also predicted a value $\log g$=1.60 $\pm$ 0.04 based on the measured diameter, available parallax, and a
    model-constrained mass.
    
    Given the current questions and indications that the models or the observations
    are not entirely consistent, we proposed to observe the star in multi-epoch
    spectrography in order to detect stellar oscillations.
    Asteroseismic analysis would provide a fresh new perspective, and
    hopefully help to unravel the current inconsistencies, while also allowing
    us to investigate the scaling relation in the non-solar regime. 
    In this paper we report the first detection 
    of stellar oscillations in HD\,122563 using the Hertzsprung SONG telescope
    located on the Observatorio del Teide.  
    This is the most metal-poor star (to our knowledge)
    to have sun-like oscillations detected\footnote{\citet{epstein2014} report
    a list of 9 metal-poor stars with detected
    oscillations with abundances larger than -2.2 dex.}.
    We measure the global seismic quantity \numax\ for HD122563 (Sect.~\ref{sec:observations})
    and along with complementary 
    information we derive a new surface gravity, radius and distance to the star (Sect.~\ref{sec:analysis}).
    New data from the Gaia mission \citep{gaia2018a} corroborate our results.
    We then discuss the 
    implications of our results on the seismic scaling relation for \logg, 
    the position of the star in the HR 
    diagram and 1D stellar models (Sect.~\ref{sec:hrdiag}). 


\section{Observations of HD\,122563}
\label{sec:observations}

\subsection{New asteroseismic observations from radial velocities.}

We obtained time series radial velocity observations with the 1-m Hertzsprung
SONG telescope equipped with an echelle spectrograph and
located at the Observatorio del Teide.  The Hertzsprung telescope is a node 
of the Stellar Observations Network Group (SONG). From April 2016  to December 
2017 we obtained an average of one spectrum per night when the object is visible.  
The spectra were reduced and calibrated using the SONG pipeline.  
Details of the Hertzsprung telescope characteristics and reduction pipeline
are given in \citet{anderson2014} and \citet{grundahl2017}. All observations were obtained
using an iodine cell for precise wavelength calibration.  A spectral resolution 
of 80\,000 and an exposure time of 900s was used throughout.

The radial velocity (RV) time series is presented in Figure~\ref{fig:timeseries}. It comprises 387 data points over a total of 
449 nights.
The typical uncertainty on the RV was found to be in the 11-14m/s range, depending on the signal-to-noise ratio in the observed spectrum.
However, as can be seen from the figure, there is a long-term trend.  We believe that this could be evidence of convection, rotation or activity.  As the trend is on the order of 300 days, far from the expected intrinsic
pulsation periods, we perform a frequency analysis directly on the 
time series produced by the pipeline.

{\subsubsection{Time-series analysis}
\label{sec:observationstime}

   \begin{figure}[h]
   \centering
   \includegraphics[width=0.5\textwidth]
{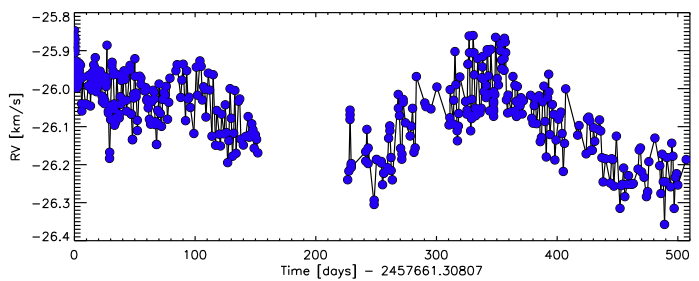}
\caption{Radial velocity time series of HD\,122563. The oscillations are on the timescale of a few days, and a long-term trend on the order of 300 days is visible.
      See Sect.~\ref{sec:observationstime}.
              }
         \label{fig:timeseries}
   \end{figure}
%

The power spectrum of the velocity time-series was initially calculated as an unweighted least-squares fit of sinusoids \citep{Frandsen95,Arentoft98,Bedding04,Kjeldsen05,Corsaro12}, and converted into power spectral density (PSD) by normalizing for the spectral resolution, namely the integral of the spectral window, of $\sim 0.06\,\mu$Hz. We also tested the case of a weighted least-squares fit to check for possible improvements in the signal-to-noise. For this purpose we used a weight assigned to each point of the radial velocity time-series according to the corresponding uncertainty estimate obtained from the SONG pipeline\footnote{It is noted in the SONG documentation that the uncertainties reported on the RV data should be 
considered with caution.}. The radial velocity uncertainties were previously rescaled in order to correct for the presence of possible outliers, following the approach presented by \cite{Butler04} (see also \citealt{Corsaro12}). Finally we measured the amplitude of the noise level in the amplitude spectrum, in the region 4-6 $\mu$Hz, outside the power excess due to oscillations, for both the weighted and unweighted case. {We found that the amplitude of the noise is lower in the un-weighted case, reaching down to $11.8$\,m~s$^{-1}$. We therefore decided to adopt the un-weighted spectrum for further analysis.}
}

   \begin{figure}[h]
   \centering
   \includegraphics[width=0.5\textwidth]
{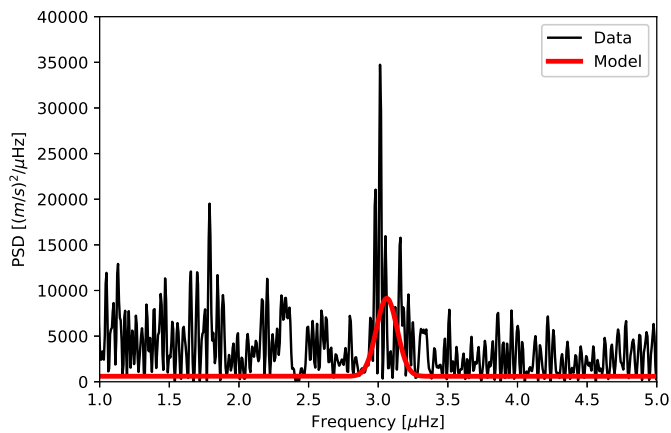}
\caption{Power spectral density and background model fit with \textsc{Diamonds} to determine \numax. The red line shows the total fit, including the oscillation power excess.
      See Sect.~\ref{sec:observationsnumax}.
              }
         \label{fig:powerspectrum}
   \end{figure}
%

{
\subsubsection{Extraction of global seismic parameters \numax}
\label{sec:observationsnumax}
We used the \textsc{Diamonds} Bayesian Inference tool (\citealt{Corsaro14}, App.~\ref{app-software}) to model the power spectral density (PSD) of the star. 
The PSD and the best-fit model are shown in Fig.~\ref{fig:powerspectrum}, and incorporates a flat noise component, {two Harvey-like profiles to account for granulation-driven signal,} and a Gaussian envelope to model the oscillation power excess \citep{Corsaro15}. A clear excess of power due to the
oscillations is detected at 3~$\mu$Hz (see Table~\ref{tab:properties}), this is referred to as \numax.
We note that the width of the power excess is narrow, and this seems to follow the trend presented by \citet{yu2018}, although their sample only goes as low as 40 $\mu$Hz.
The results reported in Table~\ref{tab:properties} 
are obtained from calculating the 16, 50, and 84 percentiles of the 
marginalised distribution of \numax. 
We note that lower and upper confidence intervals are strictly formal uncertainties, without consideration of possible systematic errors. 
{As some possible sources of errors, we also determined \numax\ 1) after filtering for the low-frequency signal, and 2) using a flat background in the power spectrum.  In both cases our results are consistent with those reported in Table~\ref{tab:properties}.}

   \begin{table}
      \caption[]{Observed Properties of HD\,122563 used in this work}
         \label{tab:properties}
         \vspace{-0.5cm}
     $$ 
         \begin{array}{p{0.1\linewidth}lllll}
            \hline\hline
            \noalign{\smallskip}
            Property & & {\rm Value} & {\rm Source}      \\
            \noalign{\smallskip}
            \hline
            \noalign{\smallskip}
			\numax\ & [\mu\mbox{Hz}] & 3.07^{+0.05}_{-0.04} & {\rm this\,work}\\
            $\theta_{\rm LD,A}$ & [\mbox{mas}] & 0.940 \pm 0.011 & \mbox{\refcre}\\
			$F_{\rm bol,A}$ &  [\mbox{erg}^{-1}\mbox{s}^{-1}\mbox{cm}^{-2}] & 13.16 \pm 0.36 \mbox{ e--8}& \mbox{\refcre} \\
			\teff$_A$ & [\mbox{K}] & 4598 \pm 41 & \mbox{\refcre} \\
            $\theta_{\rm LD,B}$ &[\mbox{mas}]& 0.926 \pm 0.011 & \mbox{\refiva}\\
			$F_{\rm bol,B}$ & [\mbox{erg}^{-1}\mbox{s}^{-1}\mbox{cm}^{-2}] & 13.20 \pm 0.29 \mbox{ e--8} & \mbox{\refiva} \\
			\teff$_B$ &[\mbox{K}]& 4636 \pm 36 & \mbox{\refiva} \\
			$\pi_{\rm HIPP}$ & [\mbox{mas}] & 4.22 \pm 0.36 & \mbox{\citet{hipp07}}\\
        	$\pi_{\rm GDR2}$ & [\mbox{mas}] & 3.444 \pm 0.063 & \mbox{\citet{gaia2018b}}\\
            \noalign{\smallskip}
            \hline\hline
         \end{array}
     $$ 
   \end{table}
%

\subsection{Literature values of the effective temperature of HD\,122563}
\label{sec:teff}
The \teff\ of HD\,122563 has been derived using many independent
methods, with all methods showing agreement within a range of $\pm$100K around 4600K. 
A recent compilation of spectroscopically derived \teff\ is given in \cite{heiter15} who recommend $4587 \pm\ 60$ K. 
\citet{cas14} used the (quasi-) model independent Infra Red Flux Method (IRFM) and 
derived a \teff\ = 4600 $\pm$ 47 K.   
Two recent analyses by C12 and \citet{iva18} (\refiva\ hereafter) using interferometry obtain results also in agreement, 4598 $\pm$ 41 K and 4636 $\pm$ 36 K.  These values have been obtained using independent determinations of the angular diameter $\theta$ and the bolometric flux  $F_{\rm bol}$  (Table~\ref{tab:properties}).  For the latter both adapt  a value of extinction\footnote{In this work we rederived $F_{\rm bol}$ using the iterative method described in \refcre, but adopting the bolometric corrections from \citet{houda2000}. This was done with the aim
of exploring the effect of a non-neglible interstellar extinction (Sect.~\ref{sec:hrdiag}).  The
new $F_{\rm bol}$ change slightly which result in 
a \teff\ = 4610/4629 K for \refcre/\refiva\ for $A_V = 0.01$ mag.}
 of $A_V = 0.01$ mag. 
 The agreement between all of the determinations provides good confidence in the \teff\ and the assumed low value of extinction.


{In this work we do not rederive \teff\ but we adopt the two independent interferometric determinations from \refcre\ and \refiva.  However, we use their reported
bolometric flux $F_{\rm bol}$ and measured angular diameter $\theta$ in order to 
correctly propagate the uncertainties on all of our inferences.  Furthermore, 
$\theta$ is used along with a parallax to perform a similar exercise in order to compare our results
(see Sect.~\ref{sect:gaia}).
By adopting both C12 and K18 one can investigate the effect of a possible source of systematic error from $\theta$, and considering an extreme value of extinction we investigate a source of error in $F_{\rm bol}$, and consequently its \teff\ (Sect.~\ref{sec:hrdiag}).}

\begin{figure}
\includegraphics[width=0.48\textwidth]{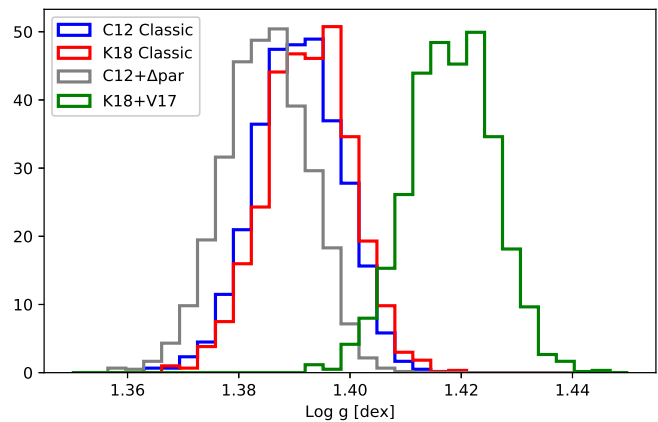}
\caption{Distributions of \logg\ derived from asteroseismic data using the seismic scaling relation.
The blue and red represent the results using the observed properties from \refcre\ and \refiva, respectively, and $f_{\nu_{\rm max}} = 1.0$.
The grey and green lines show \logg\ using the different 
solar reference values and a revised scaling relation \citep{viani2017}, respectively. See Sect.~\ref{ssec:loggnu} for details.
\label{fig:loggdistributions}}
\end{figure}

\section{Surface gravity and distance of HD\,122563}
\label{sec:analysis}
\subsection{Surface gravity and distance from asteroseismic observations\label{ssec:loggnu}}
It has been demonstrated in many papers that the surface gravity of a star can be derived
with very high precision using asteroseismic observations e.g. \citet{brown94,chaplin11,creevey12a}.
Even using simple scaling relations, such as that proposed by 
\citet{kb95} (KB95 hereafter), $\log g$ can be easily derived.  
Direct comparisons between $\log g$ derived from seismology and from other
methods have also demonstrated its accuracy \citep{mm2012,hekker2013}.  
For non-evolved stars typical errors, including systematics on the input parameters 
and accuracy, are on 
the order of 0.04 dex \citep{creevey2013}.  
However, these scaling relations are based on the Sun, and as the star begins to 
differ from the Sun, the relations may begin to deviate from solar-scaled values.
Several authors have addressed this issue recently \citep{hekker2013,coelho2015,sharma2016,viani2017,kallinger18,brogaard2018} and propose modifications to the scaling relations.  

The classical relation for the asteroseismic quantity \numax\ is 
\begin{equation}
\frac{\numax}{\nu_{\rm max\odot}} = f_{\nu_{\rm max}}\frac{g}{g_\odot} 
\sqrt{ \frac{T_{\rm eff\odot}}{\teff} } 
\label{eqn:numax}
\end{equation}
where $f_{\nu_{\rm max}} = 1.0$, $\nu_{\rm max,\odot} = 3~050 \mu$Hz 
and $T_{\rm eff,\odot} = 5~777$ K \citep{kb95}.
In this work we adopt $\log g_{\odot}$ = 4.438 dex from the IAU convention\footnote{\url{https://www.iau.org/static/resolutions/IAU2015_English.pdf} Resolution B3} 
\citep{iaub3}, 
and thus we consequently adopt $T_{\rm eff,\odot} = 5772$ K.  
Revised relations have been presented in some of the references cited above where  $f_{\nu_{\rm max}} \ne 1.00$, or $\nu_{\rm max,\odot} \ne 3~050 \mu$Hz, 
{and these are both discussed below.}

{Recent analysis by \citet{viani2017} 
(V17 hereafter)  replaces $f_{\nu_{\rm max}}$ explicitly with terms associated with the mean molecular weight and the adiabatic exponent, the theoretical basis for which has been studied in e.g. \citet{belkacem2011}.  However, as these terms can not be derived without a stellar model, we are interested in testing the classical relation, but we do consider both cases.
We also note that while these corrections can be important, they do not fully account for the effects of changes in composition on the atmospheric opacities and on the structure of the outer layers (as acknowledged by the authors).  There are important deviations between 1D and 3D stellar models in terms of the surface stratifications at very low metallicity regime and differences between 1D and 3D stratifications contribute to this surface effect \citep{tramp17}.}

\begin{figure}
\includegraphics[width=0.48\textwidth]{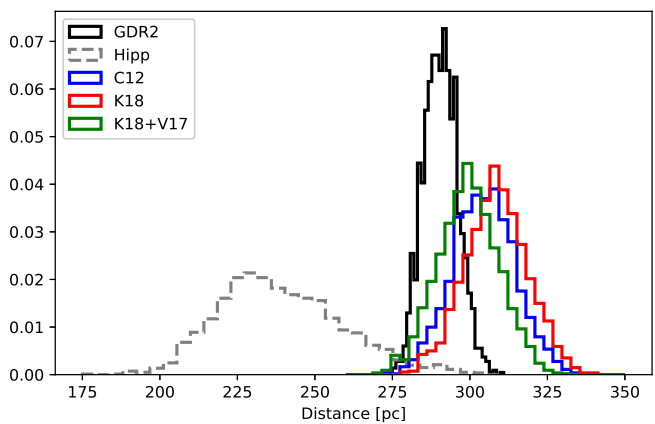}
\caption{Distances to HD\,122563 derived using parallaxes (black and grey) and asteroseismic inferences (blue, red and green)\label{fig:distance}}
\end{figure}

Using equation \ref{eqn:numax} in its classical form we calculate the surface
gravity of HD\,122563 from \numax\ and \teff. 
To correctly propagate the uncertainties reported
in \refcre\ and \refiva, and the determination of \numax, 
we performed Monte-Carlo-like simulations to derive \logg.   More specifically for each simulation we perturb
the observed quantity (\fbol, $\theta$) by adding a 
random number drawn from a Normal distribution with width 1 scaled by its 
symmetric uncertainty.  
For \numax\ we used the marginal distribution directly from Sect.~\ref{sec:observations}. 
The distributions of the resulting values of \logg\ from the simulations
are shown in Fig.~\ref{fig:loggdistributions} adopting the two referenced interferometric measurements (blue and red, respectively).  
The value of \logg$_\nu$ for \refiva\ (red) is $1.393 \pm 0.007$ dex, where the subscript $\nu$ denotes
an asteroseismically-derived value.  The median values using C12 and K18 differ by 0.0018 dex.
A possible source of systematic error arises from the definition of the solar parameters, providing a change of --0.0002 dex for \teff$_{\odot}$ and --0.0056 for \numax$_{\odot}$, and the combined change is shown for \refcre\ by the grey histogram.

{For the more recent investigations of the scaling relation, $f_{\nu_{\rm max}}$ is replaced by terms associated with the mean molecular weight $(\mu/\mu_{\odot})^{1/2}$ or a combination of $\mu$ and the adiabatic exponent $(\Gamma_1/\Gamma_{1\odot})^{1/2}$.  This is 
justified by the fact that \numax\ is expected to scale with the cut-off frequency in the atmosphere (see references above).  
In practice, these values are not readily obtained and require the use of stellar models.  For $\mu$ this may be estimated by making assumptions about the helium abundance of the star and using the observed metallicity.  For $\Gamma_1$ a stellar model is needed.  
Using the stellar models from Sect.~\ref{sec:hrdiag} we calculated $\mu_{\star} = 0.5904$ and $\Gamma_{1\star} = 1.545$, and adopted the 
solar values, $\mu_\odot = 0.6159$ and $\Gamma_{1\odot} = 5/3$, in order to derive the corrections to the scaling relation.

Applying the correction associated only with $\mu$ results in an increased $\logg$ of 0.007 dex.
However, applying the correction associated with $\Gamma_1$ has a more significant impact.  The resulting distribution for \refiva\ is represented by the green histogram in Fig.~\ref{fig:loggdistributions}. It results in a systematic change of +0.0256 dex to yield a $\logg_\nu = 1.418 \pm\ 0.007$ (see Table~\ref{tab:derivedproperties}).}

Given cosmological constraints (upper age) 
and our knowledge of stellar evolution, 
we can assume that the mass of
this evolved star is likely between 0.80 and 0.90 \msol\ (\refcre).
By adopting a conservative prior on the mass of 0.85 $\pm$ 0.05 \msol, 
the radius $R$ can be inferred from \logg\ and the mass prior.  This then
gives us access to the distance of the star, because the angular diameter has been measured.
From the classical relation with $f_{\nu_{\rm max}} = 1.0$ we derive a radius of $30.8 \pm 1.0$ \rsol\ implying a distance of $305\pm 10$ pc for \refcre\ (see Table~\ref{tab:derivedproperties}).
The smallest derived distance is $296 \pm 9$ pc using C12+V17. 
These distances are larger than that proposed by \citet{hipp07} who
measured its parallax of 4.22 $\pm$ 0.35 mas using data from the Hipparcos mission. 
The derived distances are illustrated in Fig.~\ref{fig:distance} using the same
color-code as Fig.~\ref{fig:loggdistributions}, with the latter denoted by 
the grey dashed lines.

\subsection{The surface gravity and distance from Gaia DR2 measurements\label{sect:gaia}}
The Gaia DR2 catalogue \citep{gaia2018a,gaia2018b} provides a new and more precise parallax for HD\,122563 of 3.444 $\pm$0.063 mas.
The distance to the star inferred from this newer parallax is then $290 \pm 5$ pc if we assume no prior, and thus 30\% further away than previously thought 
(see Fig.~\ref{fig:distance}, black distribution).
This value is consistent at the 1$\sigma$ level with the values obtained using asteroseismology.
Following the methodology from Sect.~\ref{ssec:loggnu}, using the 
angular diameter and the parallax measurement and adopting
a mass prior, we infer (a new radius) and \logg\ for HD\,122563.
This results in \logg$_{\rm GDR2} = 1.432^{+0.030}_{-0.033}$ and $1.445^{+0.031}_{-0.033}$ using \refcre\ and \refiva, respectively.  
We furthermore use \fbol\ to derive luminosity \lum\ (Sect.~\ref{sec:hrdiag}).
The distance $d$, radius, surface gravity and \lum\ are summarized in Table~\ref{tab:derivedproperties} under the heading 'GDR2 Parallax'.

{A potential systematic error of +0.029 mas on the parallax has been documented in the Gaia Second Data Release \citep{luri2018}.  This zeropoint corresponds to the difference between the 
median value of the quasars observed by Gaia, which are assumed to have no parallactic motion, and zero.
For completeness, we apply this
error also in our analysis, and calculate the corresponding parameters (Table~\ref{tab:derivedproperties}).   A larger parallax implies a smaller distance and radius, and
a higher value of \logg.  This is further away from the seismic value.
However, as this zeropoint is a value derived for faint quasars, there is no reason to expect it
to apply to the brighter end of the Gaia spectrum.
}

   \begin{table*}
      \caption[]{Derived properties of HD\,122563 using asteroseismology (left) and Gaia parallaxes (right).  For the seismology parameters, we show the results using the classic scaling relation (KB95) and that given in V17, using both C12 and K18 measurements.  For the results using the GDR2 parallax we show the results without and with a systematic error $s(\pi)$ of +0.029 mas on the parallax, again for C12 and K18.}
         \label{tab:derivedproperties}
         \vspace{-0.5cm}
     $$ 
         \begin{array}{p{0.05\linewidth}llllllllllllllll}
            \hline\hline
            \noalign{\smallskip}
 && \multicolumn{4}{c}{\rm Seismology}& \multicolumn{4}{c}{\rm GDR2~Parallax}\\
 && \multicolumn{2}{c}{\rm KB95}& \multicolumn{2}{c}{\rm V17}& \multicolumn{2}{c}{{\rm s(\pi)} = 0.0}& \multicolumn{2}{c}{\rm s(\pi) \ne 0.0}\\
  & & {\rm C12} & {\rm K18} &  {\rm C12} & {\rm K18} 
   & {\rm C12} & {\rm K18} &  {\rm C12} & {\rm K18} \\
            \noalign{\smallskip}
            \hline
            \noalign{\smallskip}
\logg\ & \mbox{[dex]} & 
1.391^{+8}_{-7}
& 1.393^{+7}_{-7}
& 1.416^{+8}_{-7}
& 1.418^{+7}_{-7}

& 1.433^{+30}_{-32}
& 1.446^{+30}_{-32}
& 1.441^{+30}_{-33}
& 1.453^{+31}_{-33}
\\

$d$ & \mbox{[pc]} & 
305^{+10}_{-10} &
308^{+10}_{-10} &
296^{+9}_{-9} &
299^{+10}_{-10} &

290^{+5}_{-5} &
-- &
288^{+5}_{-5} &
-- &
\\

\rad & \mbox{[\rsol]} & 
30.8^{+10}_{-10} &
30.7^{+9}_{-9} &
29.9^{+9}_{-9} &
29.8^{+9}_{-9} &

29.4^{+6}_{-7} &
28.9^{+7}_{-6} &
29.1^{+7}_{-6} &
29.7^{+6}_{-6} \\

\lum\ & \mbox{[\lsol]} & 
381^{+26}_{-26} &
392^{+27}_{-26} &
359^{+25}_{-24} &
370^{+25}_{-24} &

346^{+16}_{-15} &
347^{+15}_{-15} &
340^{+15}_{-16} &
341^{+14}_{-14} \\
            \noalign{\smallskip}
            \hline\hline
         \end{array}
     $$ 
   \end{table*}

\begin{figure}
\includegraphics[width = 0.5\textwidth]{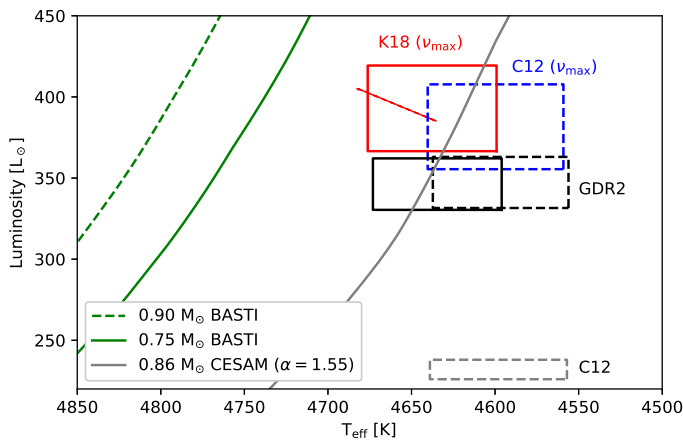}
\caption{HR diagram presenting the revised positions of HD\,122563 based on 
asteroseismic data (red and blue) and Gaia data (black) for 
\refiva\ and \refcre\ (continuous and dashed lines, respectively).  
The value presented in \refcre\ using the parallax from \citet{hipp07}
is shown as the grey dashed box.  
The red vector represents a potential shift in the median values of
\lum\ and \teff\ if we consider an extinction A$_V$ = 0.08 mag (the reference value is 0.01 mag).
Classical evolution tracks for a 0.75 and 0.90 \msol\ model from BASTI \citep{basti04} are shown by the green 
lines, while a fine-tuned 0.86 \msol\ model using the CESAM2K evolution code \citep{morel97}
with a reduced value of the mixing-length parameter compared to the solar one is shown by the grey continuous curve.
See Sect.~\ref{sec:hrdiag} for details.}
 \label{fig:hrdiag}
\end{figure}

\section{Observational constraints in the HR diagram}
\label{sec:hrdiag}
We plot the position of the new observational constraints in the HR diagram in Fig.~\ref{fig:hrdiag} using the results from the classical scaling relation and considering no systematic error on the parallax. 
We also show BASTI evolutionary tracks \citep{basti04} 
for a 0.75 and 0.90 \msol\ star 
using standard solar scaled physics  (non-canonical, alpha-enhanced shifts the tracks to 
hotter temperatures).   These tracks are frequently used in the literature.   
The new HR diagram constraints are shown using the same color-code as Fig.~\ref{fig:loggdistributions} with the grey dashed lines representing \refcre\ 
constraints.

The discrepancy between the observational error box and the evolution 
tracks already reduces from $>300$K (\refcre) to the order of
100K (considering the uncertainties) using the new constraints. 
However, assuming these models to be correct, we would still require 
\lum $\sim$550 \lsol, assuming
the \teff\ is correct.  
Given that we have two independent measures of the distance
suggesting similar luminosities it is likely that it is the stellar models that need to be adjusted, based on the assumed $A_V = 0.01$ mag (and therefore \teff).

We investigate the effect of assuming non-neglible interstellar extinction\footnote{\url{
http://stilism.obspm.fr/reddening?frame=icrs&vlong=210.63268943&ulong=deg&vlat=09.68609665&ulat=deg&valid=
}} of $A_V = 0.08$ mag by considering the maximum values from \citep{stilism},  although most studies indicate that this should not be the case.  This would increase  \fbol\ and consequently increase \teff, for a fixed $\theta$. The increase in the 
median values of \teff\ and \lum\ as a result of imposing $A_V = 0.08$ mag is indicated by the red vector in Fig.~\ref{fig:hrdiag} for \refiva.
This could explain some of the discrepancy with evolution models, although such a strong absorption seems unlikely and would also bring the various \teff\ determinations into disagreement.

In this work we have assumed a conservative mass of 0.85 $\pm$ 0.05 \msol. However, using stellar models we can further constrain the mass by
assuming a limited age range to be consistent with a Pop II star.
We performed computations of evolutionary tracks with the CESAM2k code \citep{morel97} 
assuming [Z/X] = $-2.4$ {which includes an assumed [$\alpha$/Fe] $\sim$ +0.25}, 
similar to those described in \refcre.
In order to match the constraints, we are required to lower the mixing-length parameter by $\sim 0.35$ compared to the solar one, to {$1.55 \pm 0.03$}.
Only models with masses in the range of {0.85 -- 0.87} \msol\ reach the \lum\ at an age between 10 and 12 Gyr.
A new representative stellar model (0.86 \msol, $\alpha = 1.55$) calculated from CESAM2k is shown in grey in Fig.~\ref{fig:hrdiag}. 

\section{Discussion}
\label{sec:discussion}

We determined an asteroseismic \logg\ value of $1.39 \pm 0.01$ and $1.42 \pm 0.01$ by 
using the \numax\ seismic scaling relation without and with corrections for the 
mean molecular weight and adiabatic exponent.  These values are in statistical 
agreement with those
derived using a parallax from GDR2 ($1.43 \pm 0.03$).  A recent determination from the APOGEE
survey (DR15, \citealt{sdssdr15}) 
yields a {\it calibrated\footnote{The APOGEE stellar parameters provide calibrated and uncalibrated stellar parameters, where the calibration is done independently of the other stellar parameter.  For surface gravity, the calibrated value is obtained by adding a constant that is derived from seismic calibrations of spectra.  For \teff\ however this is obtained by adding a constant derived from photometrical calibrations with models.  For the \teff, the calibrated value is not valid in our case.}} \logg\ of 1.43 dex. 
These agreements imply that the seismic scaling relation for surface gravity even without corrections is valid in the evolved, metal-poor regime.  Our work also validates the use of the 
corrective terms, although some discrepancy still remains.  
This has an important consequence, because 
today we have access to thousands of giant stars which show oscillation signatures such as \numax, and
this implies that seismic constraints in spectroscopic analyses can be safely used.  
Fixing this parameter allows more precise determinations of other spectroscopic parameters, and
can help to uncover systematic errors associated with these analyses.

A recent analysis of the CNO abundances of HD\,122563 
using 3D hydrodynamical atmospheres by \citet{collet18} 
found that their imposed surface gravity of
$1.61 \pm 0.07$ resulted in discrepant oxygen abundances between molecular and atomic 
species.  They suggested that a downward revision on the order of 0.3 dex would relieve 
this tension.   In this work, the surface gravity has been revised downward by 0.2 dex, close
to that proposed by the authors.   


Using stellar models and the constraints from the revised luminosity, we refined the mass to within 0.85 and 0.87 \msol.
Nevertheless, we still find differences compared to standard stellar models.  
We looked into the possibility of having a possible increase in interstellar extinction.  This would bring the error box in the HR diagram closer to the BASTI stellar evolution tracks (red vector in Fig.~\ref{fig:hrdiag}), because the \teff\ would increase.   However, then all different methodologies (spectroscopic, interferometric, IRFM) would no longer be in agreement.  A more recent measurement of the \teff\ with APOGEE spectra yields an uncalibrated (i.e. derived from the spectra) \teff\ of 4594 K, also in agreement with the literature \teff\ presented here.
An alternative is to increase the \lum\ only, i.e. a star that is much further away.  From this work we have two independent determinations of the distance yielding similar results.  This would also be unlikely.  
This leaves the only possibility to investigate the stellar models.  

In order to match the observational constraints with stellar models, we could increase the 
metallicity by about 1.0 dex, however, this has been measured by many different authors and 
this is a very  unlikely solution.  We could also decrease the initial helium abundance to an extremely low value, but this would be inconsistent with predictions of the primordial abundances, e.g. \citet{tytler2001}.
One of the solutions of this discrepancy is to lower the mixing-length parameter $\alpha$.  This 
decreases the \teff, without influencing the \lum.  
We estimated a shift of $\alpha$ of --0.35 compared to the solar-calibrated one by calculating refined stellar models. Such
a value is in agreement with estimations from 3D simulations.  \citealt{magic2015} suggest a reduction of $\sim$0.2 compared to the solar value for a star of this $g$, [M/H], and \teff\ (their Fig.~3), where the reduction is primarily due to the lower value of \logg.  

More recently, \citet{tayar2017}, \citet{viani2018} and \citet{creevey2017} investigated the mixing-length parameter empirically by studying samples of stars.   For the second two, their analysis  concentrated on main sequence stars and subgiants, with the lowest metallicity values close to --0.60 dex, so their results are not applicable to HD\,122563.  However, they do find a relation that depends on \logg, \teff\ and metallicity.
\citet{tayar2017} also restricted their studies to stars with [Fe/H] $>$ --1.0 dex, but looked at 
the specific case of a sample of 3,000 giants, with the aim of studying the metallicity dependence.  They found that a correction to $\alpha$ on the order of 0.2 per dex is needed.  If we apply this correction, we would require a reduction of $\sim 0.5$ dex in $\alpha$ compared to 
the solar one, not far from what we find.  
Both empirical and model results indeed support that $\alpha$ needs to be modified in the 1D stellar models.   This then would support our new parameter determinations, which consequently validates the seismic scaling relation for \logg\ in the metal-poor giant regime.

\section{Conclusions}
\label{sec:conclusions}

 In this work we described the first detection of oscillations in the metal-poor
giant HD\,122563. We determined its surface gravity using the detected \numax\ along with 
scaling relations.  By comparing with the value derived using a Gaia parallax we validated the 
classical ($f_{\nu {\rm max}} = 1.0 $) seismic scaling relations in such a metal-poor
and non-solar regime.  We found a non-significant difference of 0.04 dex.  
While these relations are valid, applying the corrections for mean molecular weight and
the adiabatic exponent results in a smaller discrepancy with the surface gravity 
derived from the Gaia parallax (0.02 dex).  

We derived  updated surface gravity, radius, and luminosity for HD\,122563.  These new 
parameters are quite different from previous determinations.  These permit us to make a new
estimate of the mass by using stellar models.  Only models with masses between 
0.85 and 0.87 \msol\ satisfy the new constraints.  
The updated luminosity along with the literature \teff\ provide a new error box in the HR diagram.
The large difference on the \teff-axis found in earlier works has significantly reduced, to the order of 100 K.  This final difference can be rectified by modifying the mixing-length parameter used in the models.  We needed a change of --0.35 compared to the solar-calibrated value, a value in agreement with 3D simulations and empirically derived values.

SONG radial velocity observations are continuing in order to determine \deltanu\ 
and resolve the individual frequencies.  These observations will help to test scaling
relations for the radius and the mass using \deltanu\ outside of the solar regime, and bring 
important constraints on the knowledge of the fundamental parameters of
this star including its age.  
Complementing these data with high precision parallax measurements will allow us to 
derive accurate masses and radii independent of models.


\begin{acknowledgements}
We are grateful for the Programme National de Physique Stellaire for financial support for this research project. This work is based on observations made with the Hertzsprung SONG telescope operated on the Spanish Observatorio del Teide on the island of Tenerife by the Aarhus and Copenhagen Universities and by the Instituto de Astrofísica de Canarias. Support for the construction of the Hertzsprung SONG Telescope from the Instituto
de Astrofisica de Canarias, the Villum Foundation, the Carlsberg Foundation and
the Independent Research Fund Denmark is gratefully acknowledged.  E.C. is funded by the European Union’s Horizon 2020 research and innovation program under the Marie Sklodowska-Curie grant agreement No. 664931.
D.S. acknowledges the financial support from the CNES GOLF grant and the Observatoire de la C\^ote d'Azur for support during his stays.
We also deeply acknowledge the Gaia team for their huge efforts in bringing us high quality data.

\end{acknowledgements}

\appendix
\section{Software and Observations \label{app-software}}
This work made use of the following free public data and software
\begin{itemize}
\item The radial velocity data were obtained using the Hertzsprung SONG telescope, which is operated on the Spanish Observatorio del Teide on the island of Tenerife by the Aarhus and Copenhagen Universities and by the Instituto de Astrofísica de Canarias.
\item Parallaxes from the ESA Gaia Space Mission Data Release 2.
\item The VizieR catalogue access tool, CDS, Strasbourg, France. The original description of the VizieR service was published in A\&AS 143, 23.
\item The radial velocities were analysed using the SONG reduction pipeline.
\item Extinction map tools from the {\tt Stilism} project \url{https://stilism.obspm.fr/}
\item The frequency analysis was done using the {\sc Diamonds} code.  This code is available at {\tt https://github.com/EnricoCorsaro/DIAMONDS}.
\item This article was prepared using {\tt overleaf}.
\item The figures were prepared using {\tt Jupyter-Notebook.}
\end{itemize}

\bibliographystyle{aa}
\bibliography{hd122}

\end{document}